\documentclass[amsmath,prb,twocolumn,showpacs,superscriptaddress]{revtex4}

\usepackage{color}
\usepackage{amsfonts}
\usepackage{graphicx}
\usepackage{pstricks}

\begin{document}

\title{Light-induced current in molecular junctions: Local field and non-Markov effects}
\date{\today}
\author{Boris D. Fainberg}
\affiliation{Faculty of Sciences, Holon Institute of Technology, Holon 58102, Israel}
\author{Maxim Sukharev}
\affiliation{Department of Applied Sciences and Mathematics,
Arizona State University at the Polytechnic Campus, Mesa, AZ 85212, USA}
\author{Tae-Ho Park}
\affiliation{Department of Chemistry \& Biochemistry, University of California
at San Diego, La Jolla, CA 92093, USA}
\author{Michael Galperin}
\affiliation{Department of Chemistry \& Biochemistry, University of California
at San Diego, La Jolla, CA 92093, USA}

\begin{abstract}
We consider a  two-level system coupled to contacts as a model  for charge pump under external 
laser pulse. The model represents a  charge-transfer molecule in a junction, and is a generalization
of previously published results [B. D. Fainberg, M. Jouravlev, and A. Nitzan. Phys. Rev. B \textbf{76},
245329 (2007)]. Effects of local field for realistic junction geometry and non-Markov response of the 
molecule are taken into account within finite-difference time-domain (FDTD) and on-the-contour 
equation-of-motion (EOM) formulations, respectively. Our numerical simulations are compared to
previously published results.
\end{abstract}


\maketitle
\section{Introduction}\label{intro}
Driven transport and coherent control at the nanoscale are well established areas of research. 
Quantum ratchets,\cite{HanggiNM11,Kemerink2011} molecular charge,\cite{RothNM10} 
spin\cite{KowenhovenRMP07,WernsdorferNM08} and heat pumps,\cite{Nitzan2007,Dlott2007} 
and nano-plasmonics\cite{HalasNL10} are just several examples of areas of recent developments.
Advances in optical techniques, in particular near-field optical microscopy, allow single
molecule manipulation\cite{Hildner2011} and induction of bond specific 
chemistry.\cite{SeidemanACR99}
Combined with molecular junction fabrication techniques,\cite{NatelsonNL07}
optical spectroscopy methods are becoming an important observation and diagnostic tool
in molecular electronics.\cite{NatelsonNL08,Ioffe2008,NatelsonNN11}

Experimental developments led to surge of theoretical activity in the field of optically assisted
transport\cite{HanggiCP04,GalperinNitzanRatnerPRL06,CuevasPRB07,CuevasPRB08}  
and optical response of molecular junctions.\cite{GalperinNitzanPRL05,GalperinNitzanJCP06,Harbola2006b,GalperinTretiakJCP08,Sukharev2010,GalperinRatnerNitzanNL09,GalperinRatnerNitzanJCP09}

In particular, Ref.~\onlinecite{GalperinNitzanRatnerPRL06} considered molecular junctions
composed of molecules with strong charge-transfer transition into their excited 
state\cite{PonderMathies1983,ColvinAlivisatos1992,SmirnovBraun1998}
as a possible constituent for light-induced molecular charge pump, when change of molecular 
dipole occurs along the junction axis. Consideration was done within a two-level (HOMO-LUMO)
model with ground and excited (HOMO and LUMO) states of the molecule strongly coupled to
different contacts. In junction setup optical excitation brings electron from occupied ground 
to empty excited state, and asymmetry in coupling to contacts assures appearance of current.
The model was treated within  non-equilibrium Green function approach, and perturbation theory 
in coupling to laser field was employed. 

Later Ref.~\onlinecite{FainbergNitzanPRB07} generalized the consideration of 
Ref.~\onlinecite{GalperinNitzanRatnerPRL06} to strong laser fields. 
Pumping optical field was treated as a classical driving force, and closed set of 
EOMs for observables (electronic populations and coherences of the levels and single time exciton 
correlation function) was formulated. One of the most  important advances in
Ref.~\onlinecite{FainbergNitzanPRB07} was consideration of chirped laser pulses,
which allowed formulation of charge transfer between ground and excited states in terms
of Landau-Zener problem. Chirped laser pulses enable to produce complete population
inversion in molecular systems (a molecular bridge) where the well-known 
$\pi$-pulse excitation\cite{All75} fails.

In realistic molecular junctions optical field driving the molecule is a {\em local field} formed by
both incident radiation and scattered response of the system (mostly plasmonic response of 
metallic contacts). Another feature of molecular junctions is hybridization of states of a molecule 
with those of contacts. The latter leads to non-Markov effects in response of the junction.

In this paper we generalize studies reported in Ref.~\onlinecite{FainbergNitzanPRB07} 
incorporating the aforementioned effects into consideration. 
Dynamics of local electromagnetic fields is simulated within the FDTD technique
for realistic geometry of a molecular junction similar to our previous 
publication.\cite{Sukharev2010}
Non-Markov effects of junction response are introduced within non-equilibrium 
Green functions equation-of-motion (NEGF-EOM) approach.

Structure of the paper is the following. After introducing the model in section~\ref{model}, 
we describe a junction geometry and numerical approach used in calculations of local 
electromagnetic fields in section~\ref{optics}. Section~\ref{cur} discusses calculation of local 
field-induced electron flux through the junction, and section~\ref{eom} introduces 
set of NEGF-EOMs. Numerical results and discussion are given in section~\ref{res}. 
Section~\ref{conclude} summarizes our findings.

\section{Model}\label{model}
A model junction consists of a molecule coupled to two metallic contacts driven by external 
radiation field. The radiation is a time-dependent local electromagnetic field $E(t)$ calculated 
within FDTD technique for bowtie geometry of the contacts (see section~\ref{optics} for details). 
Molecule is represented by a two-level system $|1>$ and $|2>$ (HOMO and LUMO 
or ground and excited states),  and is placed in a `hot spot' of the local field. 
Contacts $L$ and $R$ are assumed to be  free charge carrier reservoirs, 
each at its own equilibrium. Difference in their electrochemical
potentials defines bias applied to the junction $eV=\mu_L-\mu_R$. 
Following Refs.~\onlinecite{GalperinNitzanRatnerPRL06,FainbergNitzanPRB07} 
we consider two types of coupling between molecule and contacts: charge and energy transfer. 
Hamiltonian of the junction is
\begin{align}
 \label{H}
 \hat H(t) =& \hat H_0(t) + \hat V
\\ 
 \label{H0}
 \hat H_0(t) =& \sum_{m=1,2}\varepsilon_m \hat n_m 
 + \sum_{k\in \{L,R\}}\varepsilon_k\hat n_k
 \\
 -& \mu E(t)\left(\hat D_{12}+\hat D_{12}^\dagger\right)
 \nonumber \\
 \label{V}
 \hat V =& \sum_{m=1,2}\sum_{k\in\{L,R\}}\left(V_{km}\hat c_k^\dagger\hat d_m +\mbox{H.c.}\right)
 \nonumber \\
 +& \sum_{k_1\neq k_2\in\{L,R\}}\left(V^{en}_{k_1k_2}\hat c_{k_1}^\dagger\hat c_{k_2}\hat D_{12}
 +\mbox{H.c.}\right)
\end{align} 
Here $\hat d_m^\dagger$ ($\hat d_m$) and $\hat c_k^\dagger$ ($\hat c_k$) 
are creation (annihilation) operators of electron in level $m$ of the molecule and
in state $k$ in the contact(s), respectively, $\hat n_m=\hat d_m^\dagger\hat d_m$
is the operator of electronic population in level $m$, 
$\hat D_{12}\equiv\hat d_1^\dagger\hat d_2$ is operator of molecular de-excitation
($\hat D_{21}\equiv\hat D_{12}^\dagger$), and
$\mu$ is molecular transition dipole moment. Terms on the right-hand side of (\ref{H0})
represent molecular structure (two-level system), contacts, and coupling to the driving field.
Right-hand side of Eq.(\ref{V}) introduces electron and energy transfer 
between molecule and contact(s).  
Eqs.~(\ref{H})-(\ref{V}) introduce the model of Ref.~\onlinecite{FainbergNitzanPRB07} 
with rotating wave approximation relaxed, and with driving force treated as 
a local electromagnetic field.

\begin{figure}[t]
\centering\includegraphics[width=\linewidth]{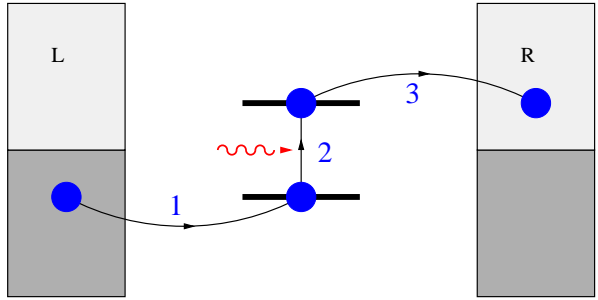}
\caption{\label{fig1}
(Color online) A sketch of local field driven molecular charge pump.
}
\end{figure}

To simulate molecules with strong charge-transfer transition with dipole moment
oriented along the junction axis below we assume that ground state (or HOMO), $|1>$,
is coupled strongly to the left contact $L$, while excited state (or LUMO) - to the right
contact $R$. Such setup works as a local field driven charge pump (see Fig.~\ref{fig1}).
Note that similar selective coupling  can also be obtained for the bridge made of 
a quantum dots as discussed in Refs.~\onlinecite{Bryllert02,Brandes08PRB,Fai10PRB}.
\section{Local field simulations}\label{optics}
Calculations of the local electromagnetic field dynamics are carried out utilizing FDTD 
technique.\cite{TafloveBook} 
Following Ref.~\onlinecite{FainbergNitzanPRB07} we assume that the incident field, 
$E_{\mbox{inc}}(t)$, has the form of a linear chirped pulse
\begin{equation}
\label{chirp}
 E_{\mbox{inc}}(t)=\mbox{Re}\left( \mathcal{E}_0 
 \exp \left( -\frac{\left(\delta^2-i\bar\mu\right)t^2 }{2}-i\omega_0t \right) \right),
\end{equation}
where $\mathcal{E}_0$ is the incident peak amplitude, $\omega_0$ is the incident frequency, and 
parameters $\delta$ and $\bar\mu$ describing incident chirped pulse are given by 
\begin{align}
 \delta^2 =& \frac{2\tau_0^2}{\tau_0^4+4\Phi''{}^2(\omega_0)}, \\
 \bar\mu =& -\frac{4\Phi''(\omega_0)}{\tau_0^4+4\Phi''{}^2(\omega_0)},
\end{align}
with $\tau_0\equiv t_{p0}/\sqrt{2\log{2}}$ (the value of the pulse duration $t_{p0}$ 
of the corresponding transform-limited pulse used in 
simulations is $9.34$~fs)  and $\Phi''(\omega_0)$ is 
the chirp rate in the frequency domain. 
Throughout the simulations the incident field is taken in the form of (\ref{chirp}) 
and is normalized to preserve the total energy of a laser pulse at different chirp rates 
according to
\begin{equation}
 \label{norm}
 \int_{-\infty}^{+\infty}dt E_{inc}^2(t)=\mbox{const}.
\end{equation}

\begin{figure}[t]
\centering\includegraphics[width=\linewidth]{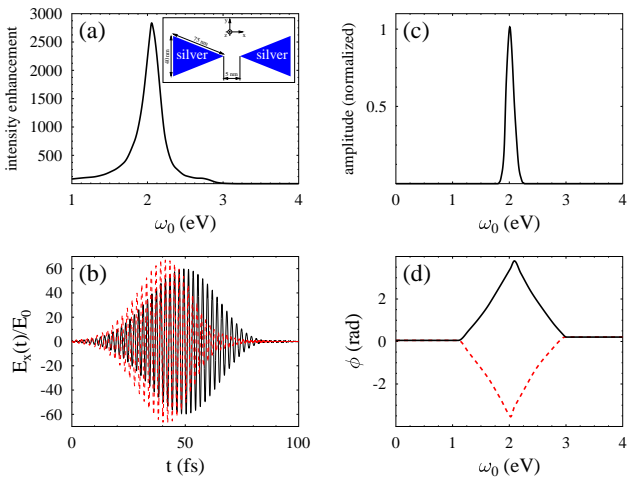}
\caption{\label{fig2}
(Color online) Results of FDTD simulations for chirped pulses exciting the bowtie antenna 
schematically depicted in the inset of panel (a). Panel (a) shows the enhancement of the local 
intensity $\left| \vec E\right|^2$ detected in the gap of the bowtie antenna as a function of the 
incident frequency. Panel (b) presents local field component of electric field along the axis of 
symmetry of the structure, $E_x$, as a function of time for two chirp rates at the plasmon resonance, 
$\omega_0=2.057$ eV: black solid line - $\Phi''(\nu_0)=-3000$ fs$^2$, red dashed line 
- $\Phi''(\nu_0)=3000$ fs$^2$. Panel (c) shows the amplitude of the local field in the frequency 
domain (note that it is independent from the phase rate). Panel (d) represents the phase of the local 
field in the frequency domain at two chirped rates: black solid line 
- $\Phi''(\nu_0)=-3000$ fs$^2$, red dashed line - $\Phi''(\nu_0)=3000$ fs${}^2$.}
\end{figure}

The geometry considered is depicted in the inset of Fig. {\ref{fig2}}a showing the top view of the 
bowtie antenna. To investigate the influence of chirped incident pulses on plasmon dynamics we 
choose incident field in the form (\ref{chirp}) and vary $\Phi''(\nu_0)=4\pi^2\Phi''(\omega_0)$. 
Below we shall write $\Phi''$, having in mind $\Phi''(\nu_0)$.
We further presume that 
the incident pulse is $x$-polarized and propagates along the $z$ axis with the incident frequency 
at the plasmon resonance (see the inset of Fig. {\ref{fig2}}a). Material dispersion of silver is taken in 
the Drude form with other numerical parameters as in Ref.~\onlinecite{Sukharev2010}. For a given 
set of material and geometric parameters the local electric field enhancement exhibits well 
pronounced plasmon resonance as seen in Fig. {\ref{fig2}}a reaching the value of 2800 near 
$2$~eV.

Our goal is to take plasmonic effects (local field enhancement and phase accumulation) directly 
into account and investigate how such crafted local fields affect transport properties of molecular 
junctions placed in the gap of bowtie antennas. However it is informative first to examine general 
features of chirped pulses interacting with plasmonic materials. It has been noted in several 
papers \cite{StockmanBergman2004,Gray2005,Brixner2007,Gray2010} that local field 
enhancement depends sensitively on the sign of chirped excitation pulses. Moreover careful 
examination of spatiotemporal dependence of local fields on chirp rates\cite{Gray2010} revealed a 
complex dynamics of plasmon wavepackets that are noticeably influenced by chirped laser pulses 
- one may find different local points for a given plasmonic system, where positive chirps lead to 
higher local fields and the other way around. 

Generally speaking, plasmonic materials can be considered as pulse shapers \cite{Weiner2009} 
due to high material dispersion near plasmon resonance, which induces a phase in the frequency 
domain resulting in shaping of the total electromagnetic field in time domain. This is illustrated in 
Fig. {\ref{fig2}}b-d, where one can clearly see that the positive chirp leads to the compression of the 
local field (Fig. {\ref{fig2}}b) and hence stronger field enhancement. While the field amplitude in 
the frequency domain is not affected by the chirp sign (Fig. {\ref{fig2}}c), 
obviously the phase of the field 
is significantly different for positive and negative chirp as shown in Fig. {\ref{fig2}}d. We note that 
one can not recover data obtained for negative chirp, for instance, by simply flipping the sign of the 
phase for the positive chirp. Additional phase induced by the plasmonic system, which depends on 
the sign of the chirp rate, makes this problem time irreversible.\cite{StockmanBergman2004}
\section{Current through the junction}\label{cur}
Time-dependent current through the junction under external driving 
is\cite{Jauho1994,Sukharev2010}
\begin{align}
 \label{IKt}
 I_K(t)=& -\frac{e}{\hbar}\left(
 \mbox{Tr}\left[\mathbf{\Gamma}^K\,\mathbf{\rho}(t)\right]
 \right. \\ +& \left.
 \frac{1}{\pi}\mbox{Im}\int_{-\infty}^{+\infty}dE\,f_K(E)\,
 \mbox{Tr}\left[\mathbf{\Gamma}^K\,\mathbf{G}^r(t,E)\right]\right)
 \nonumber
\end{align}
Here the trace is taken over molecular subspace, 
$\mathbf{\Gamma}^K$ is matrix of electronic decoherence due to coupling to
contact K
\begin{equation}
 \label{Gamma}
 \Gamma^K_{mm'}\equiv 2\pi\sum_{k\in K}V_{mk}\,V_{km'}\,\delta(E-\varepsilon_k)
\end{equation}
which is energy independent  in the wide-band approximation,
$f_K(E)$ is Fermi-Dirac thermal distribution in the contacts,
$\mathbf{\rho}(t)\equiv -i\mathbf{G}^{<}(t,t)$
is non-equilibrium reduced density matrix of molecular subsystem, 
$\mathbf{G}^{r,<}(t,t')$ are matrices in molecular subspace of  retarded and lesser projections 
of single-electron Green function
\begin{equation}
 \label{G}
 G_{mm'}(\tau,\tau') \equiv -i\langle T_c\,\hat d_m(\tau)\,\hat d_{m'}^\dagger(\tau')\rangle
\end{equation}
($T_c$ is contour ordering operator), and $\mathbf{G}^r(t,E)$ is the right side Fourier transform
of the retarded Green function
\begin{equation}
 \label{GrtE}
 \mathbf{G}^r(t,E) = \int_{-\infty}^{+\infty}dt'\, e^{iE(t-t')}\, \mathbf{G}^r(t,t')
\end{equation}
We are interested mostly in effectiveness of the device as a charge pump, i.e.
we will calculate {\em excess charge} transferred through the system during the laser pulse
\begin{equation}
 \label{QKt}
 Q_K(t) = \int_{-\infty}^t dt'\, \left(I_K(t')-I_K^{dc}\right)
\end{equation}
where $I_K(t)$ is defined in Eq.(\ref{IKt}) and $I_K^{dc}$ is current at bias induced 
steady-state condition, i.e. in the absence of radiation -- $E(t)=0$.
\section{Equations of motion}\label{eom}
Markov approximation employed in Ref.~\onlinecite{FainbergNitzanPRB07}
comes from consideration of time-local quantities only. This approach is sufficient
when one can neglect broadening of molecular states induced by hybridization 
with states in the contacts. In realistic molecular junctions such hybridization is 
non-negligible, since molecules are usually chemisorbed on at least one of the contacts.
Here (in addition to  local field formation) we are going to explore how 
non-Markovian effects influence characteristics of laser pulse induced  charge pumping. 

To keep non-Markov effects a time-nonlocal quantity -- 
single-particle Green function, Eq.(\ref{G}) --
is at the focus of our consideration. We employ Keldysh contour based EOM approach,
similar to the one employed in our earlier publication\cite{GalperinNitzanRatnerPRB07}
(see Appendix~\ref{appA} for derivation)
\begin{align}
 \label{EOMG1}
 &i\frac{\partial}{\partial\tau}G_{mm'}(\tau,\tau') = \delta_{m,m'}\delta(\tau,\tau')
  +\varepsilon_m G_{mm'}(\tau,\tau')
 \\ &
 -\mu E(t) G_{\bar m m'}(\tau,\tau')+\sum_{m_1}\int_c d\tau_1\Sigma_{mm_1}(\tau,\tau_1)
 G_{m_1m'}(\tau_1,\tau')
 \nonumber \\ &
 -i\sum_{k_1\neq k_2}\sum_{m_1}\left|V^{en}_{k_1k_2}\right|^2\int_c d\tau_1
 g_{k_2}(\tau,\tau_1)g_{k_1}(\tau_1,\tau)
 \nonumber \\ &\qquad\qquad\qquad\qquad\qquad\qquad\times
 \mathcal{G}_{\bar m\bar m_1,m'm_1}(\tau,\tau_1;\tau',\tau_1+) 
 \nonumber
\end{align}
Here $\bar m$ is molecular level other than $m$ (e.g. for $m=1$ $\bar m=2$),
$g_k(\tau,\tau')$ is a single-particle Green function of free electron in the contacts
\begin{equation}
 g_k(\tau,\tau') \equiv -i\left<T_c\,\hat c_k(\tau)\,\hat c_k^\dagger(\tau')\right>
\end{equation}
$\Sigma_{mm'}(\tau,\tau')\equiv\sum_{K=L,R}\Sigma^K_{mm'}(\tau,\tau')$ is
the self-energy due to coupling to contacts with
\begin{equation}
\label{SK}
\Sigma^K_{mm'}(\tau,\tau') \equiv \sum_{k\in K}V_{mk} g_k(\tau,\tau') V_{km'}
\end{equation}
and $\mathcal{G}$ is molecular subspace two-particle Green function
\begin{align}
\label{G2}
&\mathcal{G}_{m_1m_2,m_3m_4}(\tau_1,\tau_2;\tau_3,\tau_4)\equiv
\\ &
-\left<T_c\,\hat d_{m_1}(\tau_1)\,\hat d_{m_2}(\tau_2)\,\hat d^\dagger_{m_4}(\tau_4)\,
 \hat d_{m_3}^\dagger(\tau_3)\right>
 \nonumber
\end{align}
Note, in derivation of (\ref{EOMG1}) we treated the energy transfer term, Eq.(\ref{V}), 
at the second order of the perturbation theory.
 
Presence of many-body interaction does not allow to close hierarchy of equations exactly.
To make the problem tractable we employ Markov approximation
in treating energy transfer, last term on the right in (\ref{EOMG1}), and 
in writing EOM for two-particle GF (see below). These approximations are similar to those 
introduced previously in Refs.~\onlinecite{GalperinNitzanJCP06,FainbergNitzanPRB07}. 
Molecule-contact coupling in (\ref{EOMG1}) is treated exactly, 
thus introducing non-Markov effects into description. This leads to system of equations
(see Appendix~\ref{appA} for derivation)
%
%
\begin{widetext}
\begin{align}
\label{EOMGrE}
&
 i\frac{\partial}{\partial t}G^r_{mm'}(t,E) = \delta_{m,m'}+\left(\varepsilon_m-E\right) G^r_{mm'}(t,E)
 -\mu E(t) G^r_{\bar m m'}(t,E)-\frac{i}{2}\sum_{m_1=1,2} \Gamma_{mm_1}G^r_{m_1m'}(t,E)
\\
\label{EOMnm}
&
\frac{d}{dt}n_m(t) = 2(-1)^m \mu E(t)\mbox{Im}\left[p(t)\right]-\Gamma_{mm}n_m(t)
-\Gamma_{m\bar m}\mbox{Re}\left[p(t)\right]
+2\mbox{Re}\sum_{m_1}\int_{-\infty}^{+\infty}\frac{dE}{2\pi}G^r_{mm_1}(t,E)\Sigma^{<}_{m_1m}(E)
\\ & \qquad\qquad
-(-1)^m\bigg(B(\varepsilon_{21})N_M(t)-B(\varepsilon_{12})\left[n_1(t)-n_2(t)+N_M(t)\right]\bigg)
\nonumber \\
\label{EOMp}
&
\frac{d}{dt}p(t) = -i\mu E(t)\bigg(n_2(t)-n_1(t)\bigg)-i\bigg(\varepsilon_2+\varepsilon_1\bigg)p(t)
-\frac{\Gamma_{21}}{2}\bigg(n_1(t)+n_2(t)\bigg)-\frac{\Gamma_{11}+\Gamma_{22}}{2}p(t)
\\ &
+\sum_{m_1=1,2}\int_{-\infty}^{+\infty}\frac{dE}{2\pi}\bigg(
G^r_{2m_1}(t,E)\Sigma_{m_11}^{<}(E)-\Sigma_{2m_1}^{<}(E)\overset{*}{G}{}^r_{1m_1}(t,E)
\bigg) 
-i\sum_{m_1}B(\varepsilon_{m_1\bar m_1})\mbox{Im}\left[p(t)\right]
\nonumber \\
\label{EOMNM}
&
\frac{d}{dt}N_M(t) = 2\mu E(t)\mbox{Im}\left[p(t)\right]
-i\Sigma_{22}^{<}(\varepsilon_2)n_1(t)
+i\Sigma_{11}^{>}(\varepsilon_1)n_2(t)
-2i\bigg(\Sigma_{12}^{>}(\varepsilon_1)+\Sigma_{12}^{<}(\varepsilon_2)\bigg)\mbox{Re}
 \left[p(t)\right]
\\ & \qquad\qquad
-\bigg(\Gamma_{11}+\Gamma_{22}+B(\varepsilon_{21})\bigg)N_M(t)
+B(\varepsilon_{12})\bigg(n_1(t)-n_2(t)+N_M(t)\bigg)
\nonumber 
\end{align}
\end{widetext}
%
%
%
Here $\varepsilon_{m\bar m}\equiv\varepsilon_{m}-\varepsilon_{\bar m}$,
$n_m(t)\equiv\rho_{mm}(t)$ ($m=1,2$) are populations of molecular levels, 
$p(t)\equiv\rho_{21}(t)$ is molecular coherence,  
$N_M(t)\equiv\left\langle\hat D^\dagger(t)\hat D(t)\right\rangle\equiv\mathcal{G}_{12,12}(t+,t;t,t+)$
is the molecular excitation correlation function,
$\Gamma_{mm'}\equiv\sum_{K=L,R}\Gamma^K_{mm'}$
is the matrix of electronic decoherence due to electron transfer between the molecule and
contacts, 
with $\Gamma^K_{mm'}$ defined in Eq.(\ref{Gamma}), 
$\Sigma_{mm'}^{>,<}(E)=\sum_{K=L,R}\Sigma^{K>,<}_{mm'}(E)$ greater (lesser) 
projections of self-energy due to coupling to contacts with
\begin{align}
\Sigma^{K<}_{mm'}(E) \equiv & i\Gamma^K_{mm'}f_K(E) \\
\Sigma^{K>}_{mm'}(E) \equiv & -i\Gamma^K_{mm'}\left[1-f_K(E)\right]
\end{align}
and  $B(E)$ is the dissipation rate due to energy transfer
\begin{align}
 \label{B}
 B(E) \equiv& 2\pi\sum_{K=L,R}\sum_{k_1\neq k_2\in K} \left|V_{k_1k_2}^{en}\right|^2
 \delta(\varepsilon_{k_1}-\varepsilon_{k_2}+E)
  \nonumber\\ &\qquad\qquad\qquad\qquad\times
 f_K(\varepsilon_{k_1})[1-f_K(\varepsilon_{k_2})]
\end{align}
Note that in (\ref{EOMGrE}) we omitted term coming from energy transfer,
since contribution to the total retarded self-energy $\Sigma^r$
from molecule-contacts electron transfer $\sim\Gamma$ is much bigger than 
corresponding contribution from energy transfer 
$\sim B(\varepsilon_{21})n_2$ ($\sim B(\varepsilon_{21})(1-n_1)$) 
for $m=1$ ($m=2$) in a reasonable parameter 
range.\cite{GalperinNitzanJCP06,GalperinNitzanRatnerPRL06} 
EOMs~(\ref{EOMGrE})-(\ref{EOMNM})
form a closed set of time-dependent  equations to be solved simultaneously 
on energy grid starting from a steady-state initial condition corresponding 
to biased junction before the laser is switched on. 
Density matrix $\mathbf{\rho}(t)$ and retarded GF $\mathbf{G}^r(t,E)$
obtained as the solution are used in (\ref{IKt}) and (\ref{QKt}) 
to calculate time-dependent current and excess charge pumped through the junction, respectively.

In the limit of weak molecule-contact coupling $\Gamma\to 0$,  
neglecting local field and non-Markov effects, disregarding
off-diagonal terms in spectral function, and assuming rotating-wave approximation
Eqs.~(\ref{EOMGrE})-(\ref{EOMNM}) are reduced to results of 
Ref.~\onlinecite{FainbergNitzanPRB07} (see Appendix~\ref{appB} for details).


\section{Results and discussion}\label{res}
Here we present results of numerical simulations for the model (\ref{H})-(\ref{V}) with local field
formation and non-Markov effects taken into account as described above.  
Time dependent local electromagnetic field is calculated solving Maxwell's equations on a grid (see section~\ref{optics})
for metallic contacts of a bowtie geometry. 

Molecule is placed in a `hot spot' situated between the contacts, and local field plays a role
of external driving force in electronic calculations (as described in Section~\ref{eom}).
Unless stated otherwise parameters of the electronic simulations:
temperature is $300$K, molecular electronic level positions 
$\varepsilon_1=-1$eV and $\varepsilon_2=1$eV, elements of electronic decoherence matrix
are $\Gamma^L_{11}=\Gamma^R_{22}=0.1$eV, $\Gamma^L_{22}=\Gamma^R_{11}=0.01$eV,
$\Gamma^{L,R}_{12}=\Gamma^{L,R}_{21}=0$, coupling to external field 
$\mu\mathcal{E}_0=0.008$eV (after normalization (\ref{norm}) for $\Phi''=2000$~fs${}^2$; 
also below coupling to external field below is given renormalized according to (\ref{norm})
for particular $\Phi''$).
Fermi energy is taken as origin $E_F=0$, bias is applied symmetrically 
$\mu_{L,R}=E_F\pm |e|V_{sd}/2$. All calculations except Fig.~\ref{fig6} below are
done at equilibrium, $V_{sd}=0$. 
Only processes of energy relaxation on the molecule
are taken into account with $B(\varepsilon_{12})=0$ and $B(\varepsilon_{21})=0.1$eV.
Time grid is taken from the external driving field simulations. Energy grid spans region
from $-20$ to $20$eV with step $0.001$eV. Other parameters are introduced separately 
for each calculation.

\begin{figure}[t]
\centering\includegraphics[width=\linewidth]{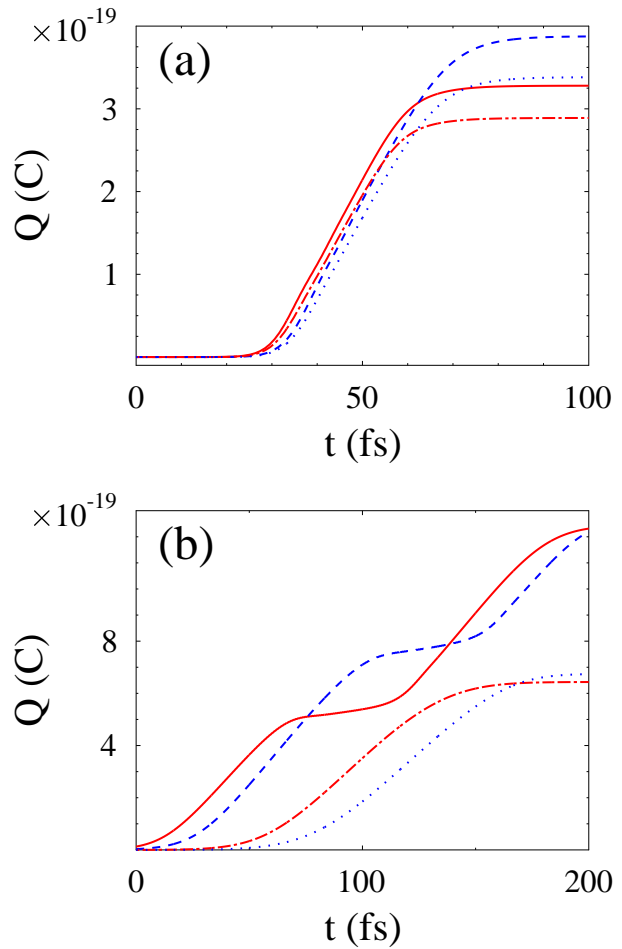}
\caption{\label{fig3}
(Color online) Time dependence of charge pumped through the junction Q(t), Eq.(\ref{QKt})
for several chirp rates.
Shown are results for (a) $\Phi''=2000$ and  $-2000$ fs${}^2$ 
without energy transfer $B(\varepsilon_{21})=0$ (solid line, red and dashed line, blue, respectively), and taking into account energy transfer term $B(\varepsilon_{21})=0.1$~eV
(dash-dotted line, red and dotted line, blue, respectively), and
(b) $\Phi''=10000$ and $-10000$ fs${}^2$ for $B(\varepsilon_{21})=0.1$~eV and
$\mu\mathcal{E}_0=0.03$ (solid line, red and dashed line, blue, respectively)
and $\mu\mathcal{E}_0=0.003$~eV (dash-dotted line, red and dotted line, blue, respectively).
}
\end{figure}

Figure~\ref{fig3} demonstrates pumped charge build-up during the laser pulse excitation. 
One sees that the local field formation leads to asymmetry in pumped charge for opposite
chirp rates. Negatively chirped incoming field creates longer local pulse (see Fig.~\ref{fig2}b),
which results in increase in total charge pumped through the junction.
Role of electron-hole excitations in the contacts on charge buildup is shown in Fig.~\ref{fig3}a.
Since processes of escape from LUMO into the right contact and energy relaxation on 
the molecule compete  for the excited state population, current (and consequently pumped charge)
diminish with increase of coupling to electron-hole excitations in the contacts.
Fig.~\ref{fig3}b shows effect of intensity of incoming pulse on the transfered charge buildup.
For higher intensity the build-up demonstrates saturation in the middle of the pulse. 
The reason for this behavior is the competition between timescales related to 
Rabi oscillation induced by local field between molecular levels
and electronic escape rate from molecule into contacts ($\sim 1/\Gamma$).
On the one hand, both negatively and positively chirped pulses  in the middle have
frequency approximately at resonance with HOMO-LUMO transition, 
$\omega\approx\varepsilon_2-\varepsilon_1$, which is a prerequisite to
effective electron transfer and thus increase in pumped charge.
On the other hand, at resonance Rabi oscillations\cite{Landau_v3_1991} 
at high enough intensities compete with electron escape rate, thus 
effectively blocking current through the junction. Depending on parameters
this may lead either to most effective charge transfer in the middle of the pulse
(dash-dotted and dotted lines in panel b), or to suppression of charge transfer
at this point (solid and dashed line in panel b).
Note that the effect is not related to non-Markov relaxation, 
i.e. this behavior is observed also in the absence
of  hybridization between molecule and contact(s) states, and its relation
to Landau-Zener problem\cite{Nitzan_2006} in terms of total charge pumped across the junction
was discussed in Ref.~\onlinecite{FainbergNitzanPRB07}. 
Note also, that with positively chirped pulse changing frequency from lower to higher 
transfered charge buildup is more effective at the start of the pulse (at lower frequencies),
while for negatively chirped pulse more effective buildup takes place at the end of the pulse
(compare solid and dashed lines in Figs.~\ref{fig3}b and \ref{fig5}b). 
Contrary to buildup suppression in the middle of the pulse, this effect is due 
to molecule-contact hybridization. The latter leads to broadening of molecular levels, and
effectiveness of  HOMO-LUMO charge transfer depends (among other conditions) 
on integral of  occupied states at HOMO and empty states at LUMO
separated by frequency of incident light $\int dE G^{<}_{11}(E) G^{>}_{22}(E+\omega)$.
Clearly, at frequencies below resonance the latter is greater than at frequencies above it.
 
\begin{figure}[htbp]
\centering\includegraphics[width=\linewidth]{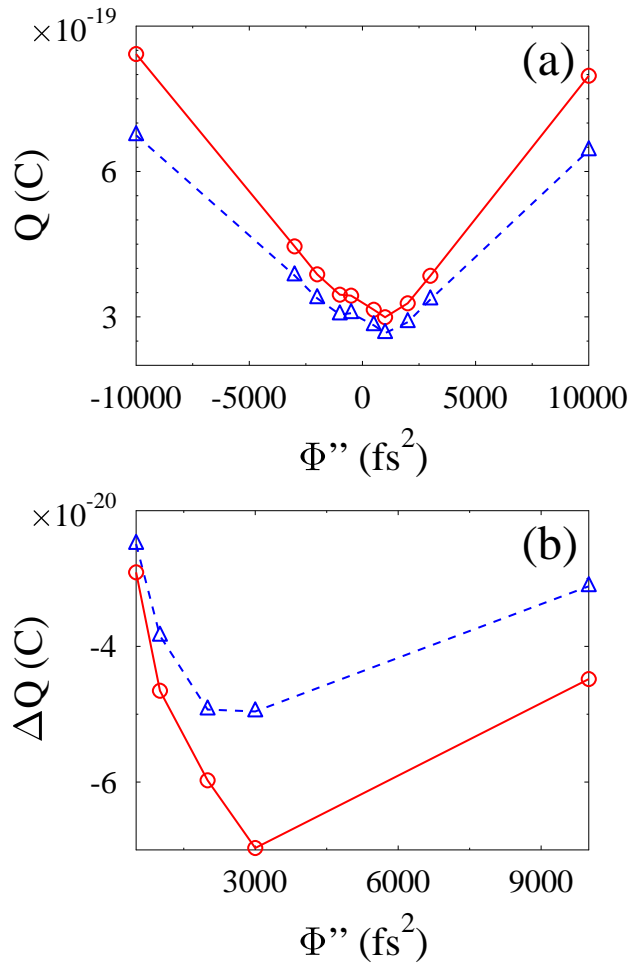}
\caption{\label{fig4}
(Color online) Charge pumped through the junction during pulse vs. chirp rate:
(a) total charge, i.e. integral of Q(t), Eq.(\ref{QKt}), over local field 
pulse duration, and (b) asymmetry in charge transfer between positively and negatively chirped 
incoming laser pulses, $\Delta Q\equiv Q(|\Phi''|)-Q(-|\Phi''|)$. 
Shown are results with ($B(\varepsilon_{21})=0.1$eV, dashed line with triangles, blue)
and without ($B(\varepsilon_{21})=0$, solid line with circles, red) energy transfer.
}
\end{figure}

Local field asymmetry relative to the sign of the chirp rate in the frequency domain
leads to asymmetry in charge pumping contrary to symmetric situation presented
 in Ref.~\onlinecite{FainbergNitzanPRB07}, as is demonstrated in  Fig.~\ref{fig4}a.
One can see that the pumped charge is almost symmetric at high rates with 
asymmetry confined to the low rate region. Difference between
charge pumped through the junction at positive and negative chirp rates is shown in 
Fig.~\ref{fig4}b. As discussed above duration of local field due to positively chirped
incoming pulse is shorter than the one due to negatively chirped analog.
This compression is the cause of less charge pumped through the system in the former case,
which results in decrease in $\Delta Q\equiv Q(\Phi'')-Q(-\Phi'')$ in the region of $\Phi''(\nu_0)$
from $0$ to $3000$ fs${}^2$.
Indeed, at the very low rates frequency of the pulse does not change much, so the asymmetry
is solely due to difference in pulse length. At a higher rates an additional factor appears:
the most effective contribution to charge transfer takes place at a particular region of frequencies 
(at and just below resonance, as is discussed above). 
This region is passed quicker in the positively chirped local pulse, and 
in the region up to $3000$~fs${}^2$ this results in increase of asymmetry,
since negatively chirped pulse spends more time in its effective frequencies zone.
Further increase of chirp rate leads to decrease and almost disappearance of the asymmetry.
The reason is decrease of ratio of the pulses difference to overall local pulse duration.

Coupling to electron-hole excitations not only diminishes pumped charge 
(compare solid and dashed lines in panel a), but also decreases asymmetry (panel b).
The latter results from the fact that rate for molecular energy relaxation (LUMO ${}\to{}$ HOMO
transition due to coupling to excitations in the contacts) is proportional to population in 
the LUMO (see discussion in Ref.~\onlinecite{GalperinNitzanJCP06}). So for higher currents also
energy relaxation will be more efficient, thus effectively compensating for the difference. 

\begin{figure}[t]
\centering\includegraphics[width=\linewidth]{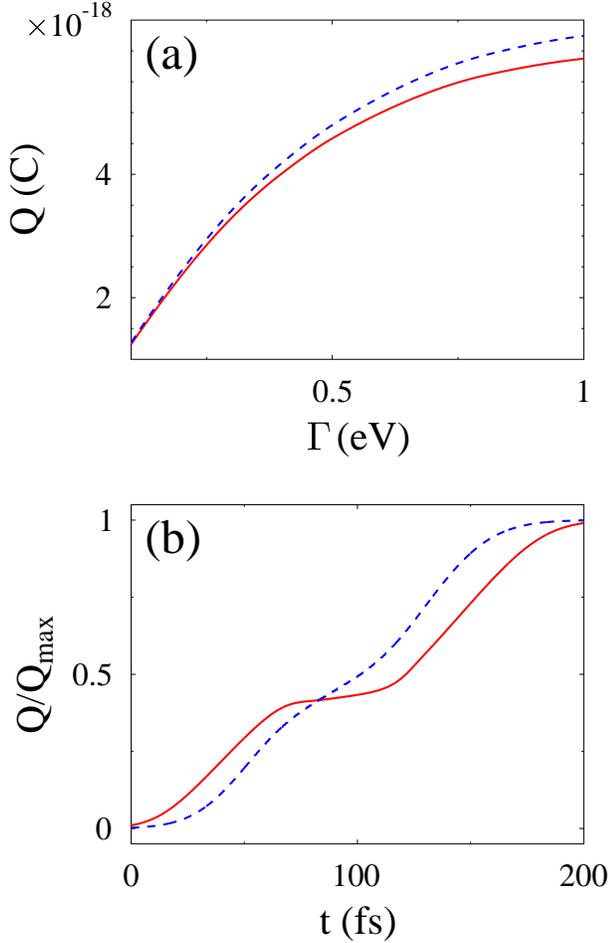}
\caption{\label{fig5}
(Color online) Dependence of charge pumping on molecule-contact states hybridization
at $\mu\mathcal{E}_0=0.03$eV.
Shown are (a) Charge pumped through the junction during pulse vs. electronic 
escape rate for chirp rate $\Phi''=10000$ fs${}^2$ (solid line, red) and $-10000$ fs${}^2$
(dashed line, blue), and 
(b) Normalized transferred charge build-up (charge normalized to total charge transferred
during the pulse) vs. time for chirp rate $\Phi''=10000$ fs${}^2$ and 
$\Gamma^L_{11}=\Gamma^R_{22}=0.1$eV (solid line, red) 
and $1$eV (dashed line, blue). 
Red line in panel (b) is the same as solid red line in Fig.~\ref{fig3}b.}
\end{figure}

Importance of non-Markov behavior for charge pump is demonstrated in Figure~\ref{fig5}.
Fig.~\ref{fig5}a shows pumped charge as function of level width (for two opposite
choices of chirp rate). 
Increase in total charge pumped through the junction with increase in hybridization 
saturates at high strength of coupling between molecule and contacts.
Such behavior is expected: at low hybridization there is only
one frequency corresponding to resonance, where pumping is most
effective, so only an 'instant' of chirped pulse contributes to charge transfer. 
As molecule-contact coupling grows the condition of resonance transition becomes less and
less strict. Eventually any frequency within the chirped pulse has roughly
same effectiveness -- this is the reason for saturation. Also, stronger coupling
means more effective molecule-contact electron transfer, which competes more effectively with
intra-molecular Rabi oscillation at resonance. This competition is demonstrated in Fig.~\ref{fig5}b,
where middle-of-the-pulse saturation (see discussion of Fig.~\ref{fig3}) disappears
for stronger molecule-contact coupling.

\begin{figure}[t]
\centering\includegraphics[width=\linewidth]{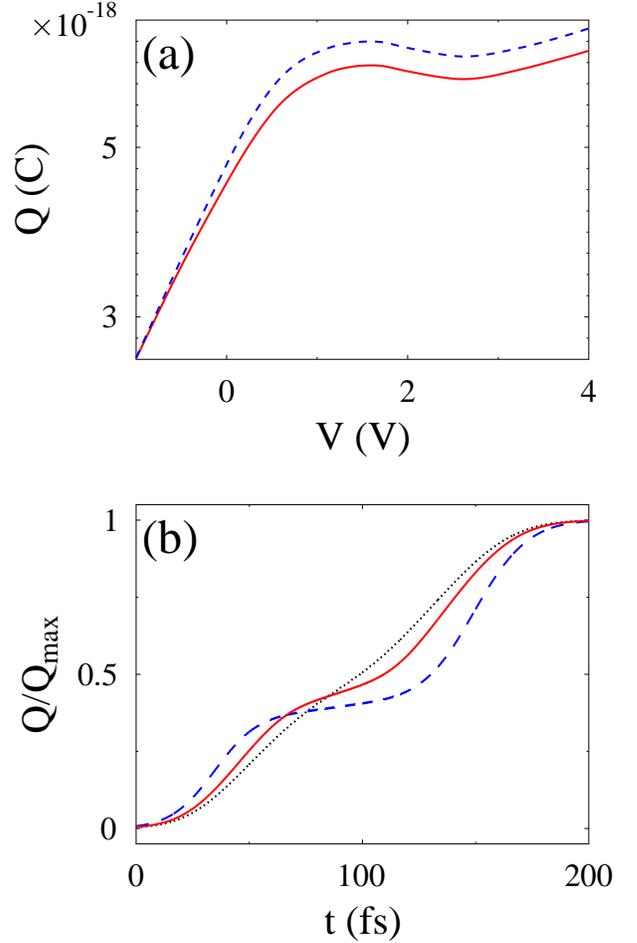}
\caption{\label{fig6}
(Color online) Dependence of charge pumping on bias.  
(a) Total excess charge pumped through the junction during pulse vs. bias 
for chirp rate $\Phi''=10000$ fs${}^2$ (solid line, red) and $-10000$ fs${}^2$
(dashed line, blue). 
(b) Excess charge build-up (normalized to total excess charge transferred
during the pulse) vs. time at chirp rate $\Phi''=10000$ fs${}^2$
for  $V_{sd}=-1$V (dashed line, blue), $0$V (solid line, red), and
$1$ V (dotted line,  black).
Here $\Gamma^L_{11}=\Gamma^R_{22}=0.5$~eV.
Red line in panel (b) is the same as solid red line in Fig.~\ref{fig3}b.
}
\end{figure}

Finally, in Figure~\ref{fig6} we discuss influence of bias on charge pumping.
Here we define optically pumped charge (excess charge) as a difference between 
charge pumped  through the junction with and without laser field.
Fig.~\ref{fig6}a demonstrates total excess pumped charge during laser pulse 
for opposite choices of chirp rate as function of bias. 
Application of bias has 2 effects on the pumping process: 
1. it depletes (populates) the HOMO (the LUMO) and 2. it may block or release
channels for electron transfer from LUMO to contact $R$.
This leads to a situations when most effective optical pumping does not
correspond to zero bias, rather we see shallow peak at $V\sim 1$V. 
Explanation is related to the fact that broadened molecular levels are
essentially a set of scattering channels with different transmission
probabilities: high conducting channels are in the center of Lorentzian, 
while channels in the sides of distribution are poor conductors.
Optical process takes electron from an occupied ground state and puts it into 
one of empty excited state. Effectiveness of the charge pump is
defined by increase or decrease of current through the junction
under optical pulse (see discussion in Ref.~\onlinecite{Sukharev2010}).
In particular, negative bias decreases effectiveness of the pump mostly due to
blocking part of LUMO-$R$ escape routes. Positive bias opens additional 
escape routes at the tail of LUMO Lorentzian, facilitating increase in pump efficiency.
However, additional effect of depleting HOMO and populating LUMO 
partially blocks optically induced HOMO-LUMO electron transfer, thus
reducing overall effectiveness of optical pumping. Competition between the two proceses
reveals itself as a shallow peak at $\sim 1$V.

Time resolved charge buildup is presented in Fig~\ref{fig6}b.  Middle-of-the-pulse
saturation observed previously at equilibrium (solid line, same as in Fig.~\ref{fig3}b) 
is enhanced  at negative (dashed line) and disappears at positive bias (dotted line).
The reason is similar to competition between Rabi frequency and escape rate discussed
above. Indeed, with negative bias partially blocking fast escape route for the electron from 
excited state into right contact, Rabi oscillation plays an important role at quasi-resonant situation
in the middle of the pulse. Positive opening additional routes makes Rabi oscillation less effective.
\section{Conclusion}\label{conclude}
We consider a two-level (HOMO-LUMO) model for optically-driven molecular charge pump.
Such pump may be realized as a junction formed by a molecule with strong charge-transfer
transition between its ground and excited states. The junction is driven by both applied bias
and laser pulse. The latter is treated as an external classical driving force.

Our consideration is the generalization of previous study\cite{FainbergNitzanPRB07}
which takes into account effects of local field (`hot spot' formation) and 
hybridization between the states of molecule and contact(s) (non-Markov effects).  
We formulate approximate closed set of EOMs for single- and two-particle GFs.
Electron transfer in the former is treated exactly.  To close set of equations the latter 
are considered within Markov approximation. Our EOMs are reduced to 
set of equation derived in  Ref.~\onlinecite{FainbergNitzanPRB07}
under several simplifying assumptions: weak molecule-contact coupling (neglect of hybridization),
neglect of non-diagonal terms in molecular spectral function, and within
rotating wave approximation.

Incoming laser pulse is assumed to be linearly chirped. Local field is calculated
within FDTD technique on a grid with bowtie antenna geometry used to represent
junction metallic contacts.
We find that contrary to symmetric behavior of the pump relative to sign
of the chirp rate, duration of the corresponding local field pulse depends on the sign of incoming chirp, which results in asymmetric operation of the pump. The asymmetry depends on
the incoming pulse chirp rate in a non-monotonic manner. Junction response to optical driving 
is symmetric to both low and high chirp rates, going through a maximum between the two extremes.
We find that this behavior is caused by the correspondence between pulse duration of 
the local field and detuning of its frequency at the end of the pulse from energy difference 
between molecular states ($\varepsilon_2-\varepsilon_1$).  

We note that at quasi-resonance
charge pump becomes ineffective, due to the competition between intra-molecular
Rabi oscillation induced by the pulse with electron transfer from molecule to contact.
Increase of the molecule-contact coupling strength  increases electron escape rate,
thus reducing ineffectiveness of the pump due to Rabi oscillations. 

Also we study the effect
of bias on optically-facilitated charge transfer through the junction. We find that in the non-Markov
situation (i.e. when hybridization between molecule and contacts is non-negligible)
most effective charge pump regime is at finite positive bias, rather than at equilibrium
as one may expect from Markov consideration of Ref.~\onlinecite{FainbergNitzanPRB07}.
The effect comes from optically-assisted charge redistribution between low and high 
conducting scattering channels in broadened molecular states, as was discussed in
our previous publication.\cite{Sukharev2010} Within the model negative bias reduces 
(positive increases) excess charge pumping due to blocking (facilitating) outgoing 
scattering channels in the excited molecular state and thus increasing (decreasing) 
the role of intra-molecular Rabi oscillations.

Finally, direct electron-hole excitation in contacts, heating,  and inelastic effects
are examples of effects beyond current consideration which may also have 
a significant impact on the properties of  a molecular charge pump.
\begin{acknowledgments}
We acknowledge support by the National Science Foundation (MG, CHE-1057930),
the US-Israel Binational Science Foundation (BF and MG, \#2008282), 
and by the UCSD (MG, startup funds).
\end{acknowledgments}
\appendix
\section{\label{appA}Derivation of Eq.(\ref{EOMG1})}
EOM for (\ref{EOMG1}) in contour variable $\tau$ starts from writing Heisenberg equation for
$\hat d_m(\tau)$
\begin{align}
 \label{Heisenberg}
 &i\frac{\partial}{\partial\tau}G_{mm'}(\tau,\tau') = \delta_{m,m'}\delta(\tau,\tau')
  +\varepsilon_m G_{mm'}(\tau,\tau')
 \nonumber \\ &
 -\mu E(t) G_{\bar m m'}(\tau,\tau')+V_{mk} G_{km'}(\tau,\tau')
 \\ &
 +\sum_{k_1\neq k_2} V^{en}_{k_1k_2}
 \mathcal{G}_{\bar m k_2,m' k_1}(\tau-,\tau;\tau',\tau+)
 \nonumber
\end{align}
The last two terms 
on the right come from electron and energy transfer terms in Eq.(\ref{V}).
Treating the two within non-crossing approximation allows to find the first exactly
within a standard procedure\cite{HaugJauho1996}
\begin{equation}
\label{T1}
 G_{km'}(\tau,\tau')=\sum_{m_1}\int_c d\tau_1 g_k(\tau,\tau_1) V_{km_1} G_{m_1m'}(\tau_1,\tau')
\end{equation}
Two particle Green function in the second term is treated (still keeping non-crossing
approximation in mind) within first order  perturbation theory  in energy transfer
\begin{align}
\label{T2}
 &\mathcal{G}_{\bar m k_2,m' k_1}(\tau,\tau;\tau',\tau) = -iV^{en}_{k_2k_1}\times
 \\ & \quad
 \sum_{m_1} \int_c d\tau_1g_{k_2}(\tau,\tau_1)g_{k_1}(\tau_1,\tau)
 \mathcal{G}_{\bar m \bar m_1,m' m_1}(\tau,\tau_1;\tau',\tau_1+)
 \nonumber 
\end{align}
Substituting (\ref{T1}) and (\ref{T2}) into (\ref{Heisenberg}) yields (\ref{EOMG1}).

Eq.(\ref{EOMGrE}) is retarded projection of Eq.(\ref{EOMG1}) with omitted energy transfer term.
The approximation is based on an estimate that in usual situation electron escape rate
should be much bigger than corresponding energy transfer, 
$\Gamma\gg B$.\cite{GalperinNitzanRatnerPRL06} 

Eqs.~(\ref{EOMnm}) and (\ref{EOMp}) is lesser projection of (\ref{EOMG1}) 
taken at equal times, $-i\mathbf{G}^{<}(t,t)$. Note that (\ref{EOMp}) is exact, while
in (\ref{EOMnm}) we employ Markov approximation in derivation of the energy transfer term
similar to previous publications,\cite{GalperinNitzanJCP06,FainbergNitzanPRB07}
for example
\begin{align}
\label{Markov}
& 
\sum_{k_1\neq k_2}\sum_{m_1}
 \int_{-\infty}^t dt_1 g_{k_2}^{>}(t-t_1)g_{k_1}^{<}(t_1-t)
 \nonumber \\ &\qquad\times
 \left\langle\hat d_m^\dagger(t)\hat d_{\bar m}(t)\hat d_{m_1}^\dagger(t_1)
 \hat d_{\bar m_1}(t_1) \right\rangle
 \nonumber \\
 &\approx\int_{-\infty}^{t}dt_1g_{k_2}^{>}(t-t_1)g_{k_1}^{<}(t_1-t)
 e^{i(\varepsilon_{m_1}-\varepsilon_{\bar m_1})(t_1-t)}
 \\
 &\qquad\times  \left\langle\hat d_m^\dagger(t)\hat d_{\bar m}(t)\hat d_{m_1}^\dagger(t)
 \hat d_{\bar m_1}(t)\right\rangle
  \nonumber \\
 &\approx 
 [1-n_{k_2}]n_{k_1}\pi\delta\left(\varepsilon_{k_2}-\varepsilon_{k_1}
 +\varepsilon_{m_1}-\varepsilon_{\bar m_1}\right)
 \nonumber\\
 &\qquad\times
 \left\langle\hat d_m^\dagger(t)\hat d_{\bar m}(t)\hat d_{m_1}^\dagger(t)
 \hat d_{\bar m_1}(t)\right\rangle
 \nonumber 
\end{align}
Using (\ref{Markov}) and similar expressions for other parts of the Keldysh contour deformed
in accordance with Langreth rules\cite{HaugJauho1996} 
in the energy transfer term of lesser projection of diagonal element 
of (\ref{EOMG1}), and utilizing (\ref{B}) leads to (\ref{EOMnm}).

Finally, Eq.(\ref{EOMNM}) is treated with Markov approximation (see Eq.(\ref{Markov}) above)
applied to both electron and energy transfer terms. Then the derivation goes along 
the lines presented in Ref~\onlinecite{FainbergNitzanPRB07}.

\section{\label{appB}Markov limit of Eqs.~(\ref{EOMGrE})-(\ref{EOMNM})}
EOMs derived in Ref.~\onlinecite{FainbergNitzanPRB07} are Markov limit 
of Eqs.(\ref{EOMnm})-(\ref{EOMNM}) within static quasiparticle approximation assumed
for molecular states. The latter implies disregarding Eq.(\ref{EOMGrE}), and 
assuming instead
\begin{equation}
i\left[G^r_{mm'}(t,E)-G^a_{mm'}(t,E)\right]=2\pi\delta_{m,m'}\delta(E-\varepsilon_m)
\end{equation} 
Then, disregarding level mixing due to coupling to contacts $\Gamma^K_{12}=\Gamma^K_{21}=0$,
Eqs.~(\ref{EOMnm}) and (\ref{EOMp}) reduce to Eqs.~(33) and (34)\footnote{Note, Markov limit
of Eq.(\ref{EOMp}) differs from Eq.(34) in Ref.~\onlinecite{FainbergNitzanPRB07}, 
since the latter is written under additional assumption of rotating-wave approximation.} 
of Ref.~\onlinecite{FainbergNitzanPRB07}. 
After omitting non-diagonal elements of self-energy
in Eq.(\ref{EOMNM}) one gets Eq.(35) of Ref.~\onlinecite{FainbergNitzanPRB07}.


\end{document}